\documentclass[]{spie}  

\usepackage{amsmath,amsfonts,amssymb}
\usepackage{graphicx}
\usepackage[colorlinks=true, allcolors=blue]{hyperref}

\title{Investigating the limited performance of a deep-learning-based SPECT denoising approach: An observer-study-based characterization}

\author[a]{Zitong Yu}
\author[b]{Md Ashequr Rahman}
\author[a,b]{Abhinav K. Jha}
\affil[a]{Department of Biomedical Engineering, Washington University in St. Louis, St. Louis, MO, USA}
\affil[b]{Mallinckrodt Institute of Radiology, Washington University in St. Louis, St. Louis, MO, USA}

\authorinfo{Further author information: (Send correspondence to Abhinav K. Jha.)\\Abhinav K. Jha: E-mail: a.jha@wustl.edu\\  Zitong Yu: E-mail: yu.zitong@wustl.edu}

\begin{document} 
This  manuscript  has  been  accepted  to  SPIE  Medical  Imaging,  2022.  Please  use  following  reference  when  citing  the manuscript.

Zitong  Yu,  Md  Ashequr  Rahman, Abhinav  K.  Jha, Investigating the limited performance of a deep-learning-based SPECT denoising approach: An observer-study-based characterization, Proc. SPIE  Medical  Imaging, 2022.
\newpage
\maketitle

\begin{abstract}
Multiple objective assessment of image-quality (OAIQ)-based studies have reported that several deep-learning (DL)-based denoising methods show limited performance on signal-detection tasks. Our goal was to investigate the reasons for this limited performance. To achieve this goal, we conducted a task-based characterization of a DL-based denoising approach for individual signal properties. We conducted this study in the context of evaluating a DL-based approach for denoising single photon-emission computed tomography (SPECT) images. The training data consisted of signals of different sizes and shapes within a clustered-lumpy background, imaged with a 2D parallel-hole-collimator SPECT system. The projections were generated at normal and 20\% low-count level, both of which were reconstructed using an ordered-subset-expectation-maximization (OSEM) algorithm. A convolutional neural network (CNN)-based denoiser was trained to process the low-count images. The performance of this CNN was characterized for five different signal sizes and four different signal-to-background ratio (SBRs) by designing each evaluation as a signal-known-exactly/background-known-statistically (SKE/BKS) signal-detection task. Performance on this task was evaluated using an anthropomorphic channelized Hotelling observer (CHO). As in previous studies, we observed that the DL-based denoising method did not improve performance on signal-detection tasks. Evaluation using the idea of observer-study-based characterization demonstrated that the DL-based denoising approach did not improve performance on the signal-detection task for any of the signal types.
Overall, these results provide new insights on the performance of the DL-based denoising approach as a function of signal size and contrast. More generally, the observer study-based characterization provides a mechanism to evaluate the sensitivity of the method to specific object properties, and may be explored as analogous to characterizations such as modulation transfer function for linear systems. Finally, this work underscores the need for objective task-based evaluation of DL-based denoising approaches.
\end{abstract}


\section{INTRODUCTION}
\label{sec:intro}

Deep-learning (DL)-based methods are showing significant interest in medical imaging, and in particular, in applications such as denoising \cite{Arabi_2021,Ramon_2020,Kaur_2018}. Typically, these methods are evaluated using figures of merits (FoMs) such as structural similarity index (SSIM) and root mean square error (RMSE)\cite{chen_2018_LEARN,Quan_2018_GAN_MRI,Ultra-low-dose_PET_GAN}. These fidelity-based FoMs measure the difference between the image obtained using DL-based approaches and a certain reference image. However, medical images are acquired for specific clinical tasks such as signal detection and quantification\cite{Barrett_OAIQ_quantum_noise_object_variability90,Barrett_OAIQ_fisher_information_95}.
Thus, ideally,  DL-based denoising approaches should be evaluated based on relevant clinical tasks.

Recently, multiple studies have shown that several DL-based denoising methods may yield limited performance on signal-detection tasks\cite{Yu575,Li_2021}. For example, it was observed that when evaluated on the task of detecting cardiac perfusion defects, a DL-based denoising approach developed for myocardial perfusion SPECT yielded similar or degraded performance as that obtained without applying denoising, even though the fidelity-based FoMs suggested that denoising was imrpoving performance\cite{Yu575}. Given the promise and wide application of DL-based denoising approaches, these studies motivate further investigation for the reasons of the limited task performance of DL-based denoising approaches.

The goal of this study is to investigate reasons for limited task performance of a commonly used DL-based denoising approach. For this purpose, we developed an approach to characterize the performance of DL-based methods. Generally, tools such as Fourier analysis, singular value decomposition\cite{Jha_SVD_2015}, modulation transfer function\cite{MTF}, and Fourier cross-talk matrix\cite{Barrett_OAIQ_fisher_information_95,Nick_FCM_2017} are used to characterize new imaging methods. However, these tools assume linearity of the underlying imaging system. In contrast, DL-based methods are typically highly non-linear and shift variant. Thus, these tools may have limited applicability in analysis of DL-based methods. An additional issue is that the characterization provided by these tools may not directly relate to task performance, although there are ongoing efforts to address this challenge. To address these issues, we propose an observer-study-based characterization of DL-based methods that quantifies the performance of these methods for specific signal properties. We then use this approach to characterize a DL-based denoising method. This characterization then provides insights on the reasons for the limited performance of this method.

The paper is organized as follows. In the Sec.~\ref{sec:Method_CNN}, we describe the details of the DL-based denoising approach. Next, we describe components of the observer-study-based characterization in Sec.~\ref{sec:observer_study_chara}. Related results are shown in Sec.~\ref{sec:results}, followed by the conclusion and discussions in Sec.~\ref{sec:conclusion}.

\section{Method}
\label{sec:method}

We conducted this study in the context of denoising SPECT images acquired at low counts. This was a simulation-based study, where objects with a circular signal within clustered-lumpy background  (CLB) and 2-D parallel-hole collimator SPECT system were simulated. A commonly used DL-based denoising approach was evaluated both on fidelity-based FoMs, including RMSE and SSIM, and on the task of detecting the signal. We then characterized the performance of the DL-based denoising approach using an observer study-based characterization. In this section, we describe the individual components of our study.

\subsection{Observer-study-based characterization}
\label{sec:observer_study_chara}
The observer-study-based characterization is rooted in principles of objective assessment of image quality, but, instead of quantifying the performance of the method over a population, quantifies performance for specific signal properties. The characterization consists of following components.

\subsubsection{Definition of task}
\label{sec:def_task}
As mentioned above, we objectively evaluated the DL-based denoising approach on the task of detecting signal in the images. We designed signals with five signal sizes and four signal-to-background ratio (SBR) values, a total of 20 types of signals, as described in more detail in the next sub-section.

We designed the signal-detection task where, in the different realizations, the signal properties (size, location, extent) were fixed while the background varied. In other words, for each signal type, we had an equivalent to  signal-known-exactly/background-known-statistically (SKE/BKS) task. Since there were 20 signal types, there were a total of 20 SKE/BKS studies. The signal-detection performance of the DL-based denoising approach was characterized by all the 20 SKE/BKS studies. 

\subsubsection{Object model}
\label{sec:obj_model}
The objects were divided into two categories, namely the signal-absent case, denoted by $H_0$, and signal-present case, denoted by $H_1$. Let $\mathbf{f}$, $\mathbf{f_b}$ and $\mathbf{f_s}$ denote the object, background, and signal, respectively. Then, the images under two hypotheses are given by
\begin{equation}
    \mathbf{f} = \begin{cases}
 &\mathbf{f_b} \text{, if } \mathbf{f} \in H_0\\ 
 &\mathbf{f_b}+\mathbf{f_s} \text{, if } \mathbf{f} \in H_1\text{.}
\end{cases}
\end{equation}

The background $\mathbf{f_b}$ was generated from a clustered lumpy background (CLB) model\cite{Bochud_1999} with parameters shown in Table \ref{tab:CLB_para}.

\begin{table}[h!]
\caption{Parameters of the CLB model used in this study.} 
\label{tab:CLB_para}
\begin{center}       
\begin{tabular}{|c|c|c|c|c|c|c|} 
\hline
Mean number of clusters & Mean number of blobs & $L_x$ [pixel]& $L_y$ [pixel]& $\alpha$ & $\beta$ & $\sigma_\phi$ [pixel] \\
\hline
150 & 20 & 5 & 2 & 1.25 & 0.5 & 24  \\
\hline 
\end{tabular}
\end{center}
\end{table}

The signal $\mathbf{f_s}$ was designed to be a circular signal located at the center of the image and characterized by a specific signal size and signal to background ratio (SBR).
The objects were pixelated into $256\times256$ pixel images with pixel size of $0.2$ cm.

We evaluated the performance of the DL-based denoising approach for five signal sizes uniformly ranging from 10 mm to 30 mm, and four SBR values varying between 1.4:1, 1.5:1, 1.8:1, and 2:1. 
We generated 200 signal-present objects and 200 signal-absent objects for each signal type for testing. One example of an object with 10 mm signal size and 1.4:1 SBR is shown in Fig.~\ref{fig:image_exp}(a).

\subsubsection{Simulating the image-formation process}
\label{sec:system_recon}

A 2D parallel-hole collimator SPECT system was simulated in this study. The system resolution of the SPECT system was 7 mm at 10 cm depth. Projections were acquired at 120 angular positions spaced uniformly over 180 degrees modeling a constant orbits. Generated projections were collapsed to 0.4 cm projection bins in a $120\times128$ projection image matrix. Then, projections were scaled to a counts level of 200,000 (referred to as normal counts) and 40,000 counts (referred to as low counts). Poisson noise was added to both the normal and low-count projections.

\begin{figure} [ht]
   \begin{center}
   \begin{tabular}{c} 
   \includegraphics[height=4.5cm]{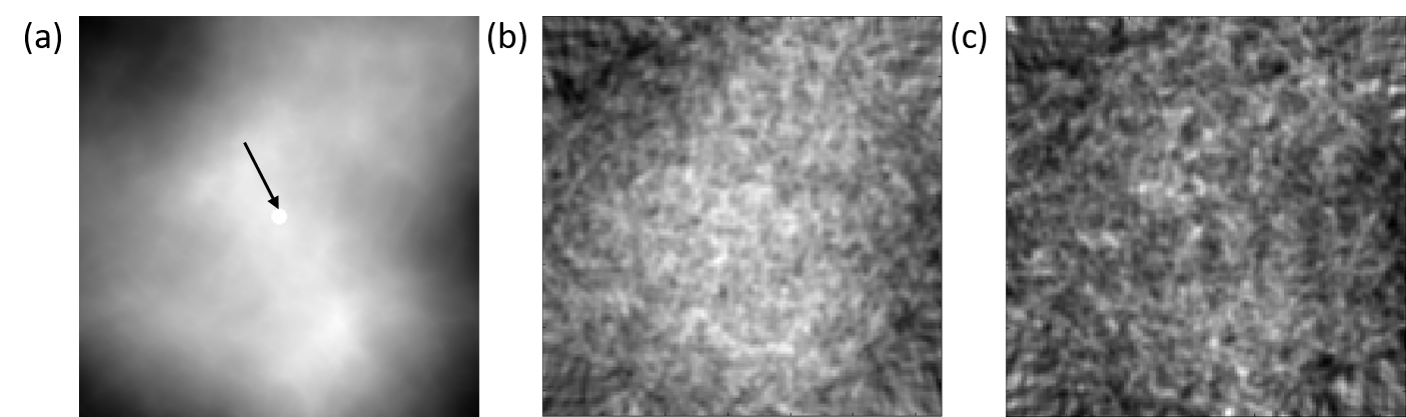}
   \end{tabular}
   \end{center}
   \caption
   { \label{fig:image_exp} (a) an example of object with 10 mm signal size and 1.4:1 SBR. (b) The same object reconstructed at normal counts level. (c)  The same object reconstructed at 20\% low counts level.}
\end{figure}

The projection data were reconstructed using an ordered-subsets expectation maximization (OSEM)-based
approach with two iterations and four subsets. For each signal type (i.e., each SKE/BKS study), we generated a total of 4,800 reconstructed images ((1,200 signal-present images $+$ 1,200 signal-absent images)$\times$ 2 counts levels). The reconstructed images were of size $128\times128$, with pixel size of 0.4 cm. Examples of reconstructed images are shown in Fig.~\ref{fig:image_exp}(b) and Fig.~\ref{fig:image_exp}(c). The low-count images were then denoised using a commonly used DL-based denoising approach described in Sec.~\ref{sec:Method_CNN}.

\subsubsection{DL-based Denoising Approach}
\label{sec:Method_CNN}

The DL-based denoising approach that we investigated\cite{Yu575} is one commonly used in medical imaging \cite{Ramon_2020,Wu_2017_CNN_CT_denoising,chen_2018_ultra_low_dose_PET,Reymann_2019_U_net_SPECT_denoise}. To develop this DL-based denoising approach, we followed best practices that have been recently outlined for developing AI-based methods for nuclear medicine \cite{Bradshaw_JNM_2021}. The approach was based on a convolutional neural network (CNN). The network has a architecture similar to a 2D U-Net\cite{Ronneberger_2015} with an encoder-decoder architecture with skip connection. The last convolutional layer was activated using a leaky ReLU activity function, yielding the output of the network.

The CNN was trained on 4,000 pairs of low-count and normal-count images, by minimizing a mean square error loss function quantifying the error between the denoised and normal-count images via the Adam optimization algorithm\cite{Adam_2015}. In the training dataset, we have included both signal-absent and signal-present images with all signal types. We conducted rigorous five-folds cross validation to ensure the network had not over-fitted. The CNN was trained using Keras with TensorFlow 1.10.0 on state-of-the-art NVIDIA V100 with 32 GB of memory. In the testing dataset, we had another 200 low-count signal-absent and 200 low-count signal-present images, for each type of signals. We refer to this CNN-based approach as the DL-based denoising approach. 

\subsubsection{Extracting task-specific information}
We evaluated the DL-based denoising approach on the task of signal detection in an observer study using an anthropomorphic model observer. 
More specifically, a channelized Hotelling observer (CHO) with anthropomorphic channels was used, where the anthropomorphic channels were six rotationally symmetric frequency channels with a starting frequency and channel width of 1/64 cycle per pixel\cite{Frey_2002_task_based}. The subsequent channels were adjacent to the previous channel and had double frequency width of the previous channel.
Channels were normalized to be orthonormal to each other. This led to a $16384\times6$ sized channel matrix.

To apply this anthropomorphic CHO, the reconstructed images were scaled to have values in the range of $\left[0,255\right]$.
The test statistic of each test image was calculated using the leave-one-out strategy\cite{Xin_Li_2017}.

\subsubsection{Figures of merit}

The test statistic of each test image was compared to a threshold, by which the image was classified into signal-present or signal-absent class. By varying the threshold, we plotted the receiver operating characteristics curve and calculated the area under the receiver operating characteristic (ROC) curve (AUC), using the LABROC4 program. 95\% confidence intervals of AUC values were also calculated. We used AUC values to quantify the performance of the DL-based denoising approach on the signal-detection task. A higher AUC value corresponds to improved performance on the signal-detection task. We compared the AUC values obtained with normal-count images, and the low-count images prior to and after applying denoising. We conducted this observer study for each signal type, yielding a characterization that directly quantifies task performance for individual signal properties. Further, we calculated the pixel-wise RMSE and SSIM between the images obtained by those two approaches and the images reconstructed at normal-count level.

\section{Results}
\label{sec:results}

The evaluation using conventional fidelity-based FoMs showed that the DL-based denoising approach provided improved performance compared to the images prior to denoising over all signal types, as shown in Table~\ref{tab:Conventional}. However, when evaluated on signal-detection task, the AUC values showed that the DL-based denoising approach did not improve performance on the signal-detection task over all signal types. In fact, the performance only worsened after applying the denoising operation. These results are similar to results in previous studies \cite{Prabhat_2021,Yang_2021,Li_2021,Zhu_2021,Yu575} and again directly contradict the observer-study-based findings.


\begin{table}[h]
\caption{The RMSE, SSIM, and AUC values associated with the images prior to
and after denoising were compared. The confidence intervals are reported with brackets around the upper and lower limits.}
\label{tab:Conventional}
\centering
\begin{tabular}{|c|c|c|c|}
\hline
&RMSE&SSIM&AUC\\
\hline
Images prior to denoising&0.0562 [0.0561,0.0563]&0.4017 [0.4007,0.4028]&\textbf{0.8050 [0.7957,0.8144]}\\
\hline
Images after denoising&\textbf{0.0302 [0.0301,0.0304]}&\textbf{0.7743 [0.7738,0.7748]}&0.6645 [0.6529,0.6762]\\
\hline
\end{tabular}
\end{table}

\begin{figure} [h]
   \begin{center}
   \begin{tabular}{c} 
   \includegraphics[height=10cm]{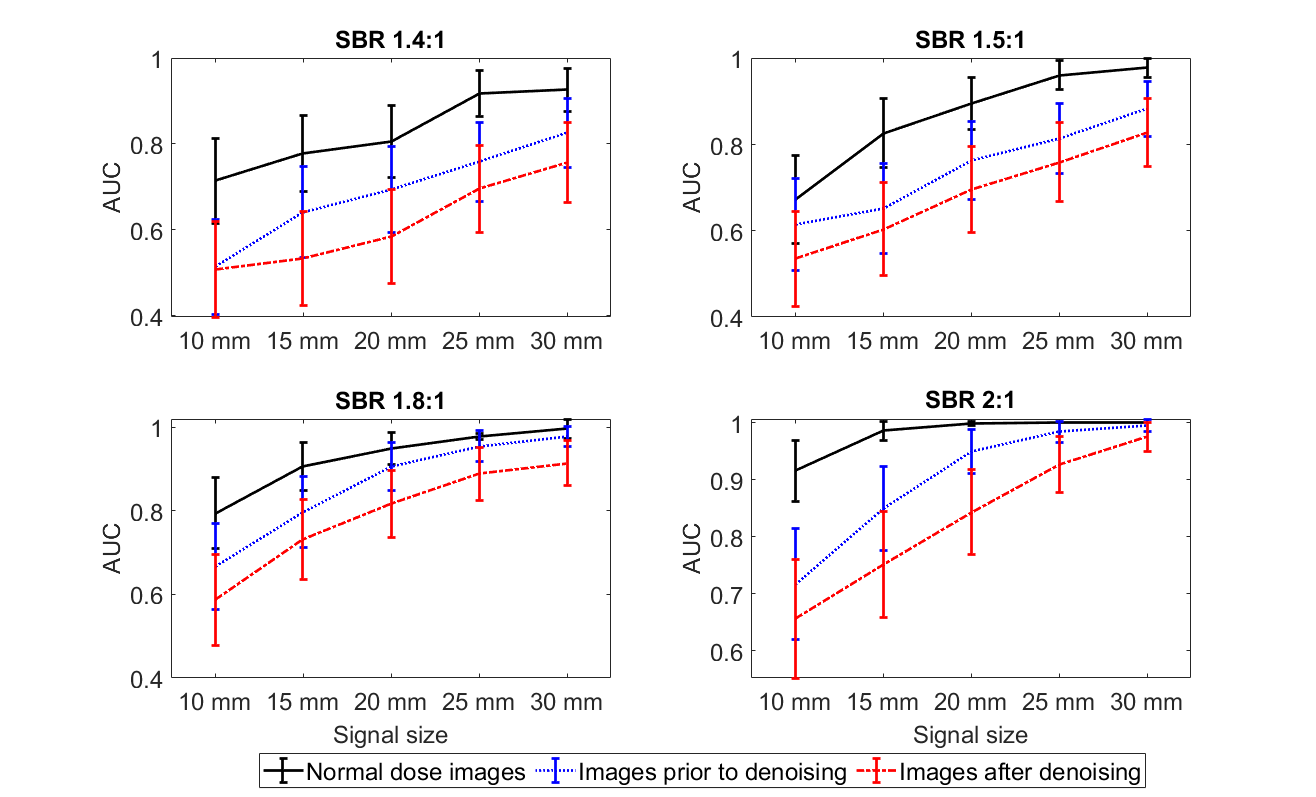}
   \end{tabular}
   \end{center}
   \caption
   { \label{fig:auc} AUC values achieved by the DL-based denoising approach, along with by normal-count images and low-count images prior to denoising for each signal type.}
\end{figure}

Given these results, we used the proposed observer-study-based characterization to investigate the limited performance of the DL-based denoising approach. Fig.~\ref{fig:auc} shows the signal-detection performance of the DL-based denoising approach evaluated via the observer study for each signal type. AUC values obtained with normal-count images and low-count images prior to denoising are also shown. We observed that performing the DL-based denoising operation does not yield superior performance on the defect-detection task for any of the signal types. This shows that the observations of limited performance of the CNN seen in previous studies \cite{Prabhat_2021,Yang_2021,Li_2021,Zhu_2021,Yu575} are valid not just for a specific population, but across a range of signal sizes and contrasts.  

Further, we observed that for an SBR of 2:1, as the size of the signal increased, the difference in the AUC values before and after applying reduced. This observation suggests that the DL-based denoising approach may be acting as a low-pass filter that suppresses high-frequency components, thus, providing poor performance when the signal size is small.

\section{Discussions and conclusion}
\label{sec:conclusion}

The manuscript demonstrates the use of an observer-study-based characterization to investigate the limited performance of a DL-based denoising approach. The characterization quantified the performance of the DL-based denoising approach for different signal types and signal-to-background ratio values on the task of defect detection. Thus, this framework may serve as a tool to investigate the performance of DL-based approaches. The underlying motivation for this study was characterizing the performance of a deep-learning-based denoising approach, given the limited performance that the approach provided on signal-detection tasks. Conventional tools to characterize imaging methods, such as Fourier analysis, singular value decomposition\cite{Jha_SVD_2015}, modulation transfer function\cite{MTF}, and Fourier cross-talk matrix\cite{Barrett_OAIQ_fisher_information_95,Nick_FCM_2017}, assume linearity and/or shift-invariance of the underlying imaging system and thus have limitations in characterizing DL-based approaches, which are  non-linear and shift variant operators. Further, these characterization tools may not directly correlate with task performance. To address these challenges, we proposed an observer-study-based characterization of DL-based methods that quantifies the performance of these methods for specific signal properties. The observer-study-based characterization provided a mechanism to evaluate the sensitivity of the method to specific object properties. Thus, this may serve as an approach to investigate the performance of DL-based methods.

A key challenge that confronts the application of DL-based methods is predicting the failure modes of the method. The proposed observer-study-based characterization may provide a mechanism to predict these failure modes. For example, an observer-study-based characterization may reveal that an AI-based denoising algorithm yields improved detection performance provided that the signal is beyond a certain size, but fails otherwise. In that case, the user of the method will be aware of the limitations of the technique prior to usage. This will lead to a more informed use of the DL-based method. 

The observer-study-based characterization may also help in learning about the working principles of the algorithm. For example, in our study, we observed that as the signal size increased, the difference in the AUC values before and after applying the denoising algorithm reduced. This suggested that the algorithm may be suppressing the high-frequency features. Learning about these working principles may help improve the interpretability of these techniques. 

In this study, it was observed here that evaluation of the DL-based method on clinical tasks was discordant with evaluation using fidelity-based figures of merit. This finding, which is similar to findings in previous studies, continues to underscore the importance of task-based evaluation. To perform such objective task-based evaluation of DL-based methods, a framework was recently outlined\cite{JHA_PET_clinics_2021493}. Further, best practices are being defined to objectively evaluate AI-based methods for nuclear medicine\cite{RELAINCE_unpublished}. Moreover, recent studies are investigating the evaluation of DL-based imaging methods for nuclear medicine on specific clinical tasks\cite{Yu_SPIE_2021,Liu_2021}. In this context, the proposed observer study-based characterization can be used to investigate the task performance for individual signal properties. For example, we evaluated a DL-based transmission-less attenuation compensation method for SPECT on the task of defect detection \cite{Yu_SPIE_2021}. The proposed observer-study-based characterization could be used to analyze the defect-detection performance of this method for different defect types. Overall, the proposed characterization provides a mechanism to evaluate the sensitivity of new imaging methods to specific object properties.

In conclusion, an observer-study-based characterization of a DL-based denoising approach enabled studying the performance of the method on the signal-detection task for specific signal properties. The characterization revealed that the limited performance of the approach on the task of defect detection was observed over a range of signal sizes and contrasts. More generally, this characterization provides a mechanism to study the performance of a broad class of DL-based approaches for medical imaging on specific clinical tasks. 

\section{Acknowledgement}
This work was supported in part by grants R21-EB024647, R01-EB031051, and R56-EB028287. The authors would like to thank Kyle Myers, PhD for helpful discussions.

\bibliography{report} 
\bibliographystyle{spiebib} 

\end{document}